\documentclass[reprint,amsmath,amssymb,pra,]{revtex4-2}
\usepackage{braket} 
\usepackage{graphicx} 
\usepackage{dcolumn} 
\usepackage{bm} 
\usepackage{color} 
\usepackage{ulem} 
\usepackage{xcolor} 
\usepackage{siunitx}
\DeclareSIUnit\gauss{G}
\usepackage{mathtools,esint}
\usepackage{blindtext}

\usepackage{hyperref}

\usepackage{tikz}
\usetikzlibrary{decorations.pathmorphing}
\usetikzlibrary{decorations.markings}
\pgfdeclaredecoration{Snake}{initial}
{
  \state{initial}[switch if less than=+.625\pgfdecorationsegmentlength to final,
                  width=+.3125\pgfdecorationsegmentlength,
                  next state=down]{
    \pgfpathmoveto{\pgfqpoint{0pt}{\pgfdecorationsegmentamplitude}}
  }
  \state{down}[switch if less than=+.8125\pgfdecorationsegmentlength to end down,
               width=+.5\pgfdecorationsegmentlength,
               next state=up]{
    \pgfpathcosine{\pgfqpoint{.25\pgfdecorationsegmentlength}{-1\pgfdecorationsegmentamplitude}}
    \pgfpathsine{\pgfqpoint{.25\pgfdecorationsegmentlength}{-1\pgfdecorationsegmentamplitude}}
  }
  \state{up}[switch if less than=+.8125\pgfdecorationsegmentlength to end up,
             width=+.5\pgfdecorationsegmentlength,
             next state=down]{
    \pgfpathcosine{\pgfqpoint{.25\pgfdecorationsegmentlength}{\pgfdecorationsegmentamplitude}}
    \pgfpathsine{\pgfqpoint{.25\pgfdecorationsegmentlength}{\pgfdecorationsegmentamplitude}}
  }
  \state{end down}[width=+.3125\pgfdecorationsegmentlength,
                   next state=final]{
     \pgfpathcosine{\pgfqpoint{.15625\pgfdecorationsegmentlength}{-.5\pgfdecorationsegmentamplitude}}
     \pgfpathsine{\pgfqpoint{.15625\pgfdecorationsegmentlength}{-.5\pgfdecorationsegmentamplitude}}
  }
  \state{end up}[width=+.3125\pgfdecorationsegmentlength,
                 next state=final]{
     \pgfpathcosine{\pgfqpoint{.15625\pgfdecorationsegmentlength}{.5\pgfdecorationsegmentamplitude}}
     \pgfpathsine{\pgfqpoint{.15625\pgfdecorationsegmentlength}{.5\pgfdecorationsegmentamplitude}}
  }
  \state{final}{\pgfpathlineto{\pgfpointdecoratedpathlast}}
}

\begin{document}

\title{
Isolated quantum-state networks in ultracold molecules
}

\author{Tom R. Hepworth}
\author{Simon L. Cornish}
\author{Philip D. Gregory}
\email{p.d.gregory@durham.ac.uk}
\affiliation{Department of Physics and Joint Quantum Centre (JQC) Durham-Newcastle, \\ Durham University, Durham, DH1 3LE, United Kingdom.}

\date{\today}

\begin{abstract}
Precise rotational-state control is at the heart of recent advances in quantum chemistry, quantum simulation, and quantum computation with bialkali molecules. Each rotational state comprises a rich manifold of hyperfine states between a given pair of rotational states, making efficient navigation challenging. Here, we describe an approach based on a heuristic and graph theory to quickly identify optimal sets of states in bialkali molecules. We find pathways to prepare the molecule in a specific state with maximum speed for any desired fidelity. We then examine networks of states where multiple couplings are present at the same time.
As example applications, we identify a closed loop of 4 states where there is minimal population leakage out of the loop during simultaneous microwave coupling; we then extend the optimisation procedure to account for decoherence induced by magnetic-field noise and obtain an optimal set of 3 states for quantum computation applications.
\end{abstract}
\maketitle

Ultracold molecules possess a rich structure of rotational and hyperfine states that provides a vast Hilbert space in which to encode quantum information. Polar molecules also allow access to long-range and anisotropic dipole-dipole interactions that can be used to engineer quantum entanglement. These properties have led to many proposed applications of ultracold polar molecules spanning the fields of quantum computation and simulation~\cite{Cornish2024}, quantum state-controlled chemistry~\cite{Karman2024}, and the precision measurement of fundamental constants~\cite{DeMille2024}.

Precise preparation of molecules in chosen quantum states is important for all applications of ultracold molecules. A broad range of experiments are now established based on ultracold bialkali molecules that are initially prepared in a single hyperfine level of their rovibronic ground state using a combination of magnetoassociation and STIRAP~\cite{Danzl2008,Ni2008,Lang2008,Takekoshi2014,Molony2014,Park2015,Guo2016,Rvachov2017,Voges2020,Stevenson2023,He2024}. Alternatively species of molecules that are amenable to laser cooling can be prepared in a single rotational and hyperfine state via optical pumping~\cite{Williams2018, Anderegg2021}. From these initially prepared states, molecules can then transferred to arbitrary rotational and hyperfine states, or superpositions of these states, using coherent microwave pulses~\cite{Ospelkaus2010,Gregory2016,Will2016,Guo2018control,Williams2018,Blackmore2020pccp}. 

A wide variety of states may be useful in experiments; the choice of states, in combination with the strength of applied magnetic or electric fields, is crucial for optimising coherence times~\cite{Blackmore2019,Gregory2021,Lin2022} and allows tuning of the strength of dipole-dipole interactions~\cite{Hermsmeier2024,Gregory2024}. Varying the prepared state can also have important effects on collisional loss~\cite{Ospelkaus2010collisions,Gregory2019,Son2020,Gregory2021collisions, Wang2021, Voges2022} and reaction dynamics~\cite{Liu2021,Hu2021}. However, the speed and fidelity with which molecules can be transferred between states is typically limited by the need to minimise off-resonant couplings to nearby unwanted states during the microwave transfer. For bialkali molecules in particular, this is further complicated by a rich and complex hyperfine structure resulting from the combination of rotational angular momentum and the angular momenta associated with the nuclear spins~\cite{Aldegunde2008,Aldegunde2017,Blackmore2023}; each rotational state comprises a dense forest of $(2I_\mathrm{A}+1)(2I_\mathrm{B}+1)(2N+1)$ hyperfine states, where $N$ is the rotational angular momentum, and $I_\mathrm{A}, I_\mathrm{B}$ are the spins associated with each nuclei. As an example, each rotational state of the $^{87}$Rb$^{133}$Cs molecule has $32(2N+1)$ hyperfine levels that are non-degenerate when a magnetic field is applied. Mixing between uncoupled-basis states can lead to many possible routes for transfer between rotational states, and the relative strengths of transitions can depend strongly on the electromagnetic fields present; a general approach to finding the best optical way to transfer from any rotational and hyperfine state to another has not yet been established.

By simultaneously driving couplings between many states of the molecule, experiments may be able to leverage the rich structure of molecules to, for example, engineer synthetic dimensions~\cite{Sundar2018, Sundar2019}. Here, internal states of the molecule can be mapped onto the sites of a lattice. Driving transitions between these internal states is analogous to molecules hopping between synthetic lattice sites. This approach has the advantage that the synthetic dimension can be easily reconfigured into many novel geometries with precise control over hopping rates and relative phases, and allows experimental exploration of systems beyond 3 dimensions~\cite{Arguello2024}. Synthetic dimensions have been used to great effect in photonic systems~\cite{Ozawa2019,Yu2025} and using the internal states~\cite{Mancini2015,Stuhl2015,Livi2016,Kanungo2022,Chen2024,Trautmann2024} or motional states~\cite{An2021,Oliver2023} of ultracold atoms for the study of novel topologies without spatial constraints. Experiments realising synthetic dimensions in molecules are yet to be reported, and one of the key challenges for molecule-based synthetic dimensions is in satisfying the requirement to form a closed network that avoids loss into unwanted states. The identification of sets of states that are well-isolated and can be described as optimal for a given experiment is nontrivial given the large number of closely-spaced molecular states available.

In this work we present a general approach to finding isolated pathways and networks comprised of the rotational and hyperfine states of bialkali molecules. We establish a simple heuristic that enables off-resonant couplings to be evaluated quickly. We then apply techniques from graph theory to search for sets of states that can be efficiently reached, and form a given geometry with minimal leakage to unwanted states. We give examples throughout by applying our approach to the $^{87}$Rb$^{133}$Cs molecule. We find a closed loop of 4 states that is accessible to experiments and compatible with 99.9\% fidelity when any of the transitions in the loop are driven with Rabi frequencies up to $\sim\qty{2}{\kilo\hertz}$, and identify an optimal set of states for realising an iSWAP quantum computing protocol~\cite{Ni2018,Picard2025}. 

\begin{figure*}
    \centering
    \includegraphics[width=1\linewidth]{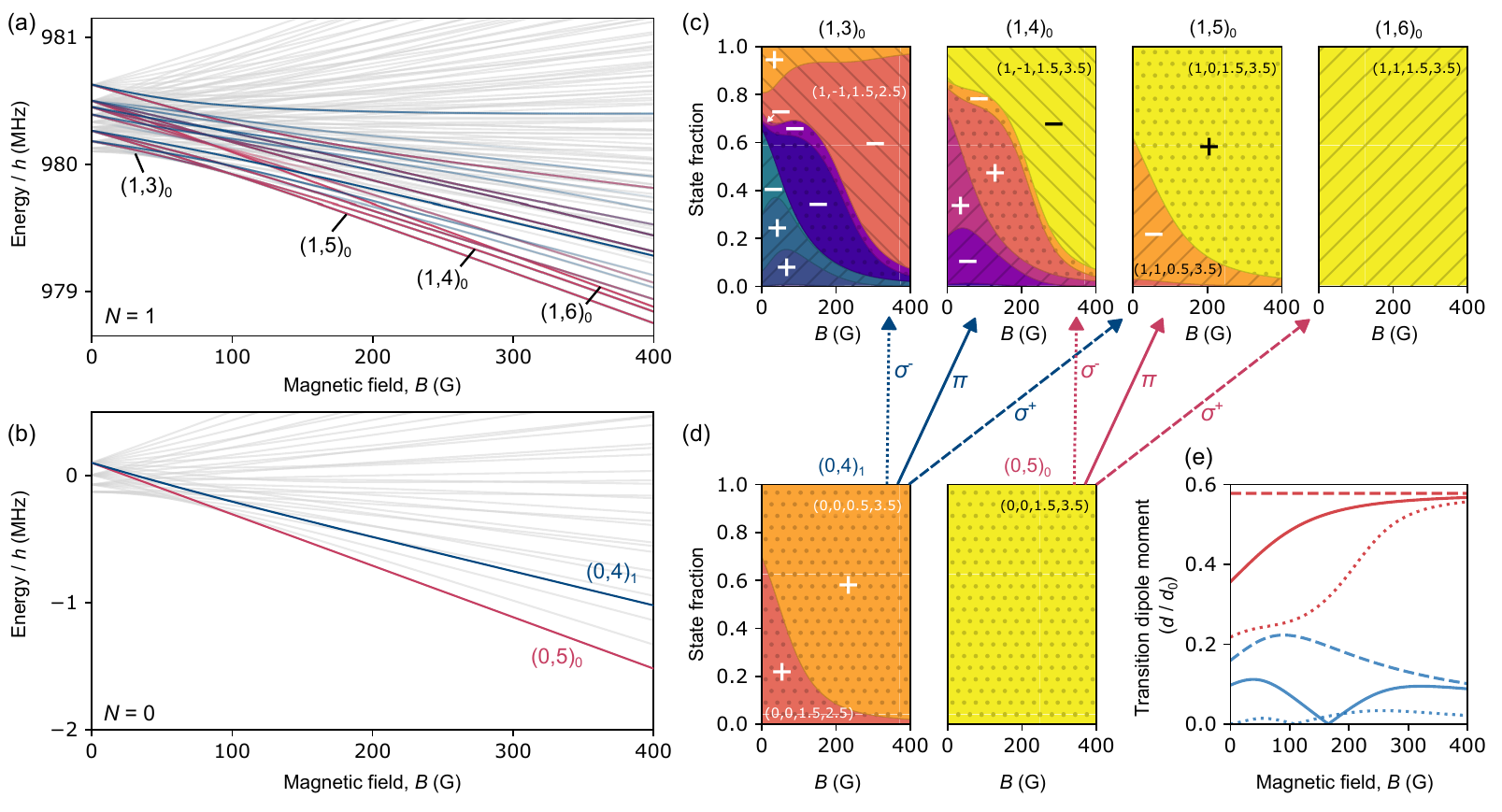}
    \caption{Rotational and hyperfine structure of $^{87}$Rb$^{133}$Cs, as an example bialkali molecule. Hyperfine Zeeman structure for (a)~$N=1$ and (b)~$N=0$ are shown. In~(b)~the $(N=0,M_F=5)_0$ and $(0,4)_1$ states are highlighted. In~(a)~the states are coloured by the strength of one-photon transitions from these states for unpolarised microwaves. Transitions from $(0,5)_0$ are red, and those from $(0,4)_1$ are blue. The lowest energy $N=1$ states with $M_F=3,4,5,6$ are labeled in (a) and their composition in the uncoupled basis given as a function of magnetic field in (c). The compositions of the initial $N=0$ states are given in (d). For all composition plots, The sign of the state coefficient is indicated by $+/-$, and hatchings represent the $M_N$ of the uncoupled component; $M_N=0$ is shown with dots, $M_N= \pm 1$ are shown with positive/negative gradient lines respectively. The largest uncoupled component at high field [and second largest for $(0,4)_1, (1,5)_0$] is labeled by $(N,M_N,m_\mathrm{Rb},m_\mathrm{Cs})$. The arrows between composition plots indicate allowed transitions with dotted, solid, and dashed lines indicating $\sigma^-,\pi$ and $\sigma^+$ transitions respectively. The magnitude of the transition dipoles associated with each of these transitions is given in (e), with red lines indicating transitions from $(0,5)_0$, blue lines from $(0,4)_1$ with the line style matching the arrows above. }
    \label{fig:zeeman-plot-and-state-comp}
\end{figure*}

\section{Results and Discussion}

\subsection{State mixing in the uncoupled basis}

We illustrate the dense structure of hyperfine states typical of bialkali molecules in Fig 1 where we plot the effect of varying the magnetic field for $^{87}$Rb$^{133}$Cs in $N=0$ and $N=1$ in Fig.~\ref{fig:zeeman-plot-and-state-comp}(a,b). At high magnetic field, the rotational angular momentum and nuclear spins decouple such that $(N, M_N, m_\mathrm{A}, m_\mathrm{B})$ becomes a good basis, where $m_\mathrm{A}, m_\mathrm{B}$ are the angular momentum projections associated with the nuclear spins. The nuclear Zeeman shifts are on the order of $\mu_\mathrm{N}=0.76$\,kHz\,G$^{-1}$, while the typical energy scale of the hyperfine structure can vary from 100\,Hz to 100\,kHz depending on the bialkali combination. There is therefore often a range of intermediate magnetic fields where the Zeeman shifts and hyperfine structure are comparable such that the uncoupled basis states are mixed. In certain species such as RbCs, experiments generally take place in this intermediate regime as high field is not easily accessible to use in experiments, requiring large magnetic fields around~$1000$\,G. At lower (non-zero) magnetic fields, the only good quantum numbers are $(N, M_F)$ where $M_F$ is the total angular momentum projection $M_F=M_N+m_\mathrm{A}+m_\mathrm{B}$. As these quantum numbers are insufficient to uniquely identify some states, we label the states as $(N, M_F)_k$ where $k$ is an index that counts up in order of increasing energy for a given $N, M_F$ at high field.

Electric dipole transitions between rotational states can be driven with microwaves following the selection rules $\Delta N = \pm 1$ and $\Delta M_N = 0, \pm1$ with the latter depending on the polarisation of the driving field. The nuclear spin projections $m_\mathrm{A}, m_\mathrm{B}$ are expected to remain unchanged during such a transition, as this would require coupling to the much weaker magnetic dipoles associated with the nuclear spin. Despite this, mixing between hyperfine levels~\cite{Aldegunde2008,Aldegunde2017} causes the polarisation-dependent selection rule to effectively become $\Delta M_F = 0, \pm1$. This results in multiple strong transitions from any one state even for polarised driving fields. We illustrate this state mixing with $^{87}$Rb$^{133}$Cs in Fig.~\ref{fig:zeeman-plot-and-state-comp}.

The strength of transitions between uncoupled basis states in $N=0$ and $N=1$ is $d_0/\sqrt{3}$ where $d_0=1.2$\,D is the molecule-frame dipole moment~\cite{Takekoshi2014, Molony2014}. The mixing of uncoupled basis states modifies these transition strengths, such that the degree of mixing determines the relative strength of transitions. In Fig.~\ref{fig:zeeman-plot-and-state-comp} we focus on transitions from the \mbox{$N=0$} states $(0,4)_1$ and $(0,5)_0$ in which molecules can be prepared directly after association~\cite{Takekoshi2014,Molony2014}. The $N=1$ states in  Fig.~\ref{fig:zeeman-plot-and-state-comp}(a) are colour coded by the relative strength of one-photon transitions from these initial states. We see that across the magnetic field range of 0-400\,G there are many transitions accessible even from just these two initial states. The strength and spacing of these transitions varies significantly with magnetic field. 

We show the composition for the chosen $N=0$ states, and the lowest energy $N=1$, $M_F=3,4,5,6$ states in Fig.~\ref{fig:zeeman-plot-and-state-comp}(c,d). The $(0,5)_0$ and $(1,6)_0$ states are spin stretched and so do not participate in the mixing, with the states well-described by the uncoupled basis at all non-zero magnetic fields. For other pairs of states, the strength of the transition depends on the overlap between components that share the same nuclear spin. In Fig.~\ref{fig:zeeman-plot-and-state-comp}(e) we show how the transition dipole moments evolve between all of the selected states. For transitions from $(0,5)_0$, as the field is increased the allowed transitions to the $N=1$ states become stronger, with each asymptotically approaching the value of $d_0/\sqrt{3}$ as the nuclear spin projections become well defined for each of the $N=1$ levels. Conversely, transitions from $(0,4)_1$ to the levels shown become weak in the high field limit as the overlap between the excited-state and ground-state components with the same nuclear spin becomes small. Note, that the transition strength associated with the $\sigma^-$ and $\pi$ transitions from $(0,4)_1$ vanishes at magnetic fields of 109\,G and 166\,G respectively. This is due to cancellation effects between state components with opposite sign.

\subsection{A model for off-resonant excitation}

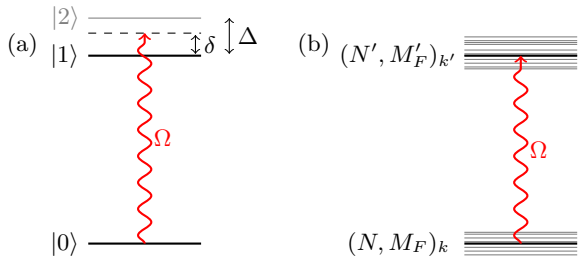
\begin{figure}
    \centering
    \begin{tikzpicture}[scale=2.5]

        \begin{scope}[xshift=-1.7cm]
            
            \node[align=left] at (0.05,1.055) {(a)};

            \draw[black, thick] (0.4,0.0) -- (1,0)node[pos=0,left]{$\ket{0}$};
    
            \draw[black, thick] (0.4,1.0) -- (1,1.0) node[pos=0,left]{$\ket{1}$};
            \draw[black, thin, dashed] (0.4,1.12) -- (1,1.12);
            \draw[gray, thin] (0.4,1.2) -- (1,1.2) node[pos=0,left]{$\ket{2}$};
            \draw[->, red, thick, decorate, decoration=snake] (0.7,0.0) -- (0.7,1.12) node[pos=0.5,right]{$\Omega$};
    
            \draw[<->] (1.15,1.01) -- (1.15,1.2) node[pos=0.5, right] {$\Delta$};
            \draw[<->] (0.97,1.01) -- (0.97,1.11) node[pos=0.5, right] {$\delta$};
    
        \end{scope}

        \begin{scope}[xshift=+1.7cm]
            \node[align=left] at (-1.8,1.055) {(b)};
                    
            \draw[gray, thin] (-1,1.1) -- (-0.4,1.1);
            \draw[gray, thin] (-1,1.03) -- (-0.4,1.03);
            \draw[gray, thin] (-1,0.96) -- (-0.4,0.96);
            \draw[gray, thin] (-1,1.06) -- (-0.4,1.06);
            \draw[gray, thin] (-1,1.07) -- (-0.4,1.07);
            \draw[gray, thin] (-1,0.98) -- (-0.4,0.98);
            \draw[gray, thin] (-1,1.00) -- (-0.4,1.00);
            \draw[gray, thin] (-1,1.01) -- (-0.4,1.01);
            \draw[gray, thin] (-1,1.08) -- (-0.4,1.08);
            \draw[gray, thin] (-1,0.94) -- (-0.4,0.94);
            \draw[gray, thin] (-1,0.93) -- (-0.4,0.93);

            \draw[gray, thin] (-1,0.06) -- (-0.4,0.06);
            \draw[gray, thin] (-1,0.03) -- (-0.4,0.03);
            \draw[gray, thin] (-1,0.05) -- (-0.4,0.05);
            \draw[gray, thin] (-1,-0.04) -- (-0.4,-0.04);
            \draw[gray, thin] (-1,0.01) -- (-0.4,0.01);
            \draw[gray, thin] (-1,-0.02) -- (-0.4,-0.02);
            \draw[gray, thin] (-1,-0.06) -- (-0.4,-0.06);
    
            \draw[black, thick] (-1,-0.0) -- (-0.4,0) node[pos=0,left]{$ $};
    
            \draw[black, thick] (-1,1.0) -- (-0.4,1.0) node[pos=0,left]{$ $};
    
    
            \draw[->, red, thick, decorate, decoration=snake] (-0.7,0.0) -- (-0.7,1) node[pos=0.5,right]{$\Omega$};
    
            \node[align=left] at (-1.35,0.0) {$(N, M_F)_k$};
            \node[align=left] at (-1.35,1.0) {$(N', M'_F)_{k'}$};

        \end{scope}

    \end{tikzpicture}
    \caption{Energy level schematics.  (a)~The 3-level model that we use as a heuristic to quickly evaluate off-resonant couplings in the many-level molecule. We consider 2 states $\ket{0}$ and $\ket{1}$ coupled resonantly with Rabi frequency~$\Omega$ and detuning~$\delta$, in the presence of a third state $\ket{2}$ that has an allowed transition from $\ket{0}$ and that is detuned from $\ket{1}$ by~$\Delta$. (b)~Each rotational level of the molecule is comprised of a manifold of densely packed hyperfine states. We wish to resonantly couple single hyperfine states occupying neighboring rotational levels (black lines) with a driving field that induces Rabi oscillations with a frequency $\Omega$ (shown in red). Off-resonant couplings to other states (gray lines) can lead to deviation from the ideal two-level system.}
    \label{fig:energylevels}
\end{figure}

\begin{figure}[t]
    \centering
    \includegraphics{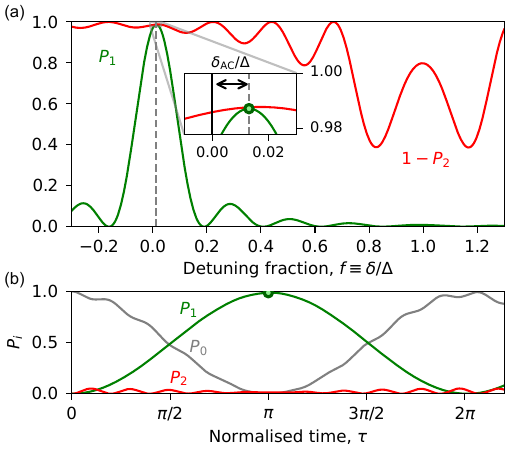}
    \caption{Analytic 3-level dynamics of populations $P_i$ for $\kappa=10$, $D=2.3$. (a) $P_1$ and $1-P_2$ against detuning fraction $f$ after evolution for time $\tau = \pi$. The inset highlights the AC-Stark shift of the transition caused by the applied driving field and the off-resonant state. (b) Populations against normalised time $\tau$, while $f=\delta_\text{AC}/\Delta$. The circular point in the inset of (a) and in (b) highlight evaluation at the same parameters.}
    \label{fig:3-lev-dynamics}
\end{figure}

To illustrate the detrimental effects of off-resonant excitation, we construct a simple 3-level model, as shown in Fig.~\ref{fig:energylevels}(a). We will later use an expression derived from this model as a simple heuristic that is sufficient to rank molecular transitions even when there are many more states available, as represented in Fig.~\ref{fig:energylevels}(b). We consider the population initialised in the state $\ket{\mathrm{0}}$, and the target is to coherently transfer the population to the final state $\ket{\mathrm{1}}$. To do this, a square-pulsed driving field is applied near-resonance with the transition from $\ket{\mathrm{0}}$ to $\ket{\mathrm{1}}$ for time $t$, detuned in energy by $\hbar\delta$. This field drives transitions from $\ket{\mathrm{0}}$ both to $\ket{\mathrm{1}}$ and to the off-resonant state $\ket{\mathrm{2}}$. The difference in energy between $\ket{\mathrm{1}}$ and $\ket{\mathrm{2}}$ is $\hbar\Delta$. We parameterise the problem in terms of dimensionless quantities: the normalised off-resonant detuning $\kappa \equiv \Delta/\Omega$, the coupling ratio $D\equiv d_{02}/d_{01}$, the detuning fraction $f\equiv\delta/\Delta$, and the normalised time $\tau\equiv t\Omega$. Here, $\Omega$, and $d_{01}$ are the Rabi frequency and transition dipole moment for the near-resonantly driven transition $\ket{\mathrm{0}}\leftrightarrow\ket{\mathrm{1}}$ and $d_{02}$ is the transition dipole moment for the off-resonant $\ket{\mathrm{0}}\leftrightarrow\ket{\mathrm{2}}$ transition. 

Under the rotating-wave approximation, with no coupling between the two excited states one can solve these dynamics analytically for the populations of the states $P_i$ after the pulse (see Supplemental Material). Figure~\ref{fig:3-lev-dynamics} shows the populations after the pulse, holding the parameters $\kappa=10$, $D=2.3$, while varying either $f$ or $\tau$. In Fig.~\ref{fig:3-lev-dynamics}(a), we show the populations after a pulse of normalised time $\tau=\pi$, for different values of $f$. The sinc profile of the excitation spectrum is apparent for $P_1$, however the presence of the third state $\ket{2}$ modifies it in two ways. Firstly, the spectrum appears shifted because as the microwave field dresses the system, the presence of the off-resonant state causes an AC-Stark shift of the initial state $\ket{0}$. The energy shift of the transition is $\hbar\delta_{\mathrm{AC}}$, where $\delta_{\mathrm{AC}}/\Delta = (\sqrt{1+(D^2/\kappa^2)}-1)/2 \approx D^2/(4\kappa^2) + \mathcal{O}(D^4)$ and the expansion is taken around $D=0$. Secondly, the amplitude is reduced due to an off-resonant excitation to the state $\ket{2}$, as the excitation spectrum to state $\ket{2}$ falls off like a sinc function away from resonance.

\begin{figure}
    \centering
    \includegraphics[width=1\linewidth]{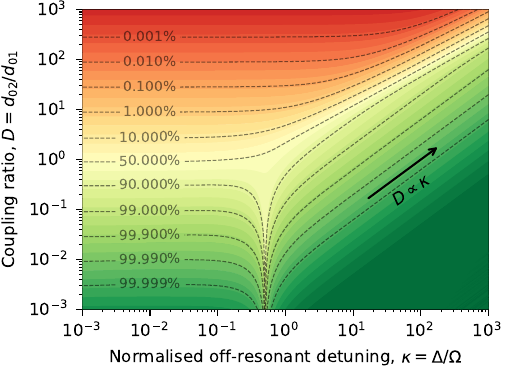}
    \caption{Phase space plot of equation $2\left< P_\mathrm{1} \right>^\text{comp}_t$, showing the average expected transfer to the intended state while compensating for the AC-Stark shift as a function of the normalised off-resonant detuning ratio~$\kappa$, and the relative coupling strength of the states~$D=d_{01}/d_{02}$. The high-fidelity contour lines when $\kappa>1$ follow $D\propto \kappa$ as is highlighted and is evident from Eq.~\ref{eqn:analytic-transfer-2-asmp-expanded}.}
    \label{fig:3-level-phase-intended}
\end{figure}

The presence of the off-resonant state is most significant when the off-resonant state detuning $\Delta$ is small compared to the Rabi frequencies $\Omega$, or the coupling to the off-resonant state is larger than the intended state. It modifies the achievable transfer fidelity for a given pulse time, or equivalently the pulse time required to reach a given fidelity. The analytic solutions to this problem contain a time-independent term that is equal to the time-averaged state occupation $\left< P_i \right>_t$. In the ideal 2-level model, evaluating this term for the target state $\ket{\mathrm{1}}$ and doubling will give $2\left< P_\mathrm{1} \right>_t=1$. For the 3-level model, this quantity is always less than 1 due to the coupling to $\ket{\mathrm{2}}$ taking some population away from $\ket{\mathrm{1}}$. Figure \ref{fig:3-level-phase-intended} shows $2\left< P_\mathrm{1} \right>_t$ (see Supplemental Material for the form) over the phase space of $D$ and $\kappa$, while compensating for the AC-Stark shift by setting $f=(\sqrt{1+(D^2/\kappa^2)}-1)/2$. We see that when $\kappa\ll1$, the off-resonant excitation depends only on $D$, whereas when $\kappa\gg1$ the off-resonant excitation depends on both $D$ and $\kappa$. There is an interesting region in Fig.~\ref{fig:3-level-phase-intended} where $\kappa=0.5$ such that $\Delta=\Omega/2$. Here, one of the two-level dressed states that results from the intended coupling is degenerate with $\ket{2}$. The $\ket{0}$ component of this dressed state can couple to state $\ket{2}$ leading to a resonant enhancement of loss from the ideal 2-level system even when $D$ is small. Crucially, our aim in this work is to find sets of states that occupy the lower-right of Fig.~\ref{fig:3-level-phase-intended} where the contour for a given fidelity has a linear relationship between $D$ and $\kappa$. 

In the limit of $\Delta\gg\Omega$, the quantity $2\left< P_\mathrm{1} \right>_t$ is approximately equal to the fidelity of the transfer, and so we use this as a proxy for optimisation. We asymptotically expand around $\kappa\rightarrow\infty$ to get
\begin{equation}\label{eqn:analytic-transfer-2-asmp-expanded}
     2\langle P_\mathrm{1}\rangle_t^{\text{comp}} \overset{\kappa \rightarrow \infty}{\approx} \left(F \equiv 1 - \frac{D^2}{4 \kappa^2}\right) + O(\kappa^{-4}),
\end{equation}
where we have defined the expected fidelity, $F$, of a transfer. A concordant result (up to a factor in the infidelity) in this limit can be found through perturbation theory~\cite{Ni2018}. Parameterising the fidelity in terms of the ``number of 9's'', which we denote $\eta$ defined by $F \equiv 1 - 10^{-\eta}$,
and rearranging this equation for the $\pi$-pulse time $t_\pi = \pi/\Omega$, gives
\begin{equation}\label{eqn:pulse-time}
    t_{\pi} = \frac{\pi}{\Delta}\biggl(\frac{D}{2}\biggr)10^{\eta/2}.
\end{equation}
This has an exponential dependence on $\eta$ with a linear coefficient dependent solely on the relative transition strengths and detuning of the off-resonant state. This coefficient may be used as a metric to compare sets of transitions without having to first set a target fidelity or Rabi frequency; smaller values allow for the highest fidelity for any given time budget, or the fastest transition speed for any required fidelity.

We note that it is possible to use shaped pulses to minimise the extent of the excitation spectrum. However, to realise a performance gain over square pulses with the same overall time, the pulse must be continuously phase modulated with a DRAG-like pulse \cite{Wilhelm2009DRAG} as the light shift become time dependent such that $\delta_{\text{AC}}\rightarrow\delta_{\text{AC}}(t)$.  
Some proposals for such schemes further engineer a spectral null at the location of the undesired state \cite{Marxer2024FastDRAG}. The use of shaped pulses increases the experimental complexity, and to our knowledge forgoes the existence of a closed form analytic solution for the population dynamics of three levels. We therefore continue our analysis assuming the use of square pulses, however we do expect that a similar approach could be employed to shaped pulses. Through simulations we have found empirically that in the limit of $\kappa\rightarrow \infty$, $t_\pi$ for a given pulse shape has the functional form $t_\pi=(\pi/\Delta)\cdot f(D)\cdot10^{\eta/m}$, where $f(D)$ is some polynomial function of $D$, and $m$ is an integer.

\subsection{Testing the performance of the heuristic}

To efficiently evaluate transitions in the many-level molecule, we use our 3-level model for expected fidelity to form a heuristic. That is, a function that is fast to compute and gives us close to the correct answer. This allows us to bypass the need to perform a slow numerical simulation of the time dynamics to find the fidelity for each possible configuration. We reduce the many-level problem to a set of 3-level subsystems. For each off-resonant level $\ket{j}$ with $j\in\{2,...,n\}$, we consider the triplet of states $\{\ket{0}, \ket{1}, \ket{j}\}$. Each of these triplets can then be modeled using the 3-level analysis shown previously, with the off-resonant state affecting the population transfer to the intended state as defined by Eq.\,\ref{eqn:analytic-transfer-2-asmp-expanded}. The overall infidelity is then approximated by the sum of all triplet infidelities. In doing this we make the assumption that the effects from any given off-resonant state are independent of the others. Rearranging for the expected time for a $\pi$-pulse on the transition $\ket{0}\leftrightarrow \ket{1}$ yields
\begin{equation}\label{eqn:pulse-time-heuristic}
    t_{\pi(0\leftrightarrow1)} = \frac{\pi}{2}\biggl({\sum_i\frac{D_i^2}{\Delta_i^2}}\biggr)^{1/2}\,10^{\eta/2}.
\end{equation}
The index $i$ iterates through all off-resonant states. 
$D_i$ is the ratio of the off-resonant coupling (from $\ket{i}$ to whichever of $\ket{0}$ or $\ket{1}$ lies in the opposite manifold) to the intended $\ket{0}\leftrightarrow\ket{1}$ coupling.
Similarly, $\Delta_i$ is the detuning of the off-resonant state $i$ from the nearest of $\ket{0}$ or $\ket{1}$. 

\begin{figure}[t]
    \centering
    \includegraphics{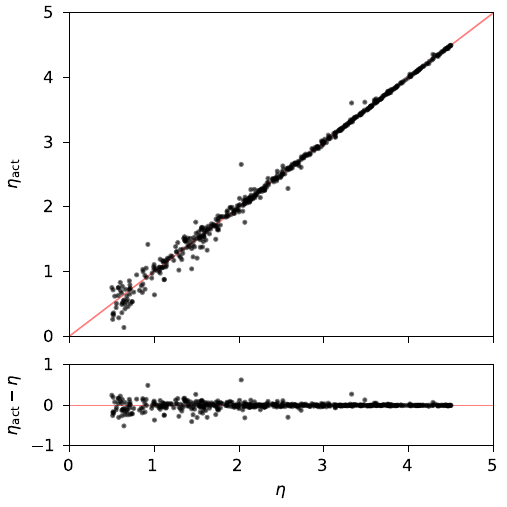}
    \caption{Deviation of the fidelity value predicted from numerical simulation to that predicted from the heuristic. The solid red line represents perfect equivalence.}
    \label{fig:heuristic-accuracy}
\end{figure}

To test the validity of our approach, we set up a system of two manifolds with randomised spacings and coupling strengths between the states. We select one state from each manifold, choose a random value of $\eta$ as our target, and calculate $t_\pi$ using Eq.~\ref{eqn:pulse-time-heuristic}. We then run a complete numerical simulation for the system dynamics with this pulse time, compensating for the AC-Stark shift, and extract the peak population transfer to get the actual number of 9's of fidelity $\eta_{\mathrm{act}}$. We repeat this process many times for various randomised manifolds and desired fidelities. The results of these simulations are shown in Fig.~\ref{fig:heuristic-accuracy}. The standard deviation of the difference between the numerical simulation and our heuristic for $\eta>2$ is $\Delta \eta = 0.05$ which corresponds to a percentage uncertainty on $t_\pi$ of $10^{\Delta \eta/2}-1 = 6.4\%$. This suggests that it is reasonable to quote predicted values of the pulse time to 2 significant figures where a relatively high fidelity is expected.

\subsection{Molecular transitions as a graph}

Using a sequence of microwave pulses, we can potentially transfer molecules between any pair of rotational and hyperfine states. However, given the highly mixed nature of states in the intermediate-field regime, it is not obvious what the best sequence of microwave pulses are. The natural structure to solve and visualise this problem is an undirected graph that has states and transitions represented as vertices and edges, respectively. The weights of each edge are defined by $t_{\pi(a\leftrightarrow b)}$ (Eq.~\ref{eqn:pulse-time-heuristic}), here for an initial and final state $a$ and $b$ of the molecule. A path on this graph, formed of a connected sequence of states, represents consecutive $\pi$ pulses driving from one state to the next. By formulating the system as a graph, we are able to use well-known graph-theoretic techniques and algorithms to solve this problem.

The coupling between the molecule transitions and the microwave field is polarisation dependent. We incorporate polarisation impurity of the microwaves into our model by splitting the electric field strength into 2 components. We assume the microwaves in the correct polarisation mode have electric field $|E|$ but also include a field $(1-p)|E|$ to account for other polarisation modes, so that $p=1$ represents perfect polarisation and $p=0$ represents unpolarised microwaves. In this case, the $\pi$-pulse duration for a given fidelity is defined by the sum over transitions
\begin{equation}
    t_{\pi(a\leftrightarrow b)} = \frac{\pi}{2}\left(\sum_{i,\text{on-pol}}\frac{D_i^2}{\Delta_i^2}+(1-p)\sum_{i,\text{off-pol}}\frac{D_i^2}{\Delta_i^2}\right)^{1/2} 10^{\eta/2}.
\end{equation}
Here the first sum is over states $\ket{i}$ that have a dipole allowed transition under the polarisation that drives the $a\leftrightarrow b$ transition. The second is over the states that have a dipole allowed transition under the other two non-ideal polarisations.
We will primarily care about the two limits $p=1$ and $p=0$, which we will call $t^{\text{pol}}_{\pi(a\leftrightarrow b)}$ and $t^{\text{unpol}}_{\pi(a\leftrightarrow b)}$ respectively. For intermediate cases $0<p<1$, a linear interpolation between the two values such that
\begin{equation}
t_{\pi(a\leftrightarrow b)} \approx p t^{\text{pol}}_{\pi(a\leftrightarrow b)} + (1-p) t^{\text{unpol}}_{\pi(a\leftrightarrow b)},
\end{equation}
is a good approximation without the need to re-evaluate the whole expression since $t_{\pi(a\leftrightarrow b)}$ is monotonically increasing with respect to $p$.

\begin{figure}
    \centering
    \includegraphics[width=\linewidth]{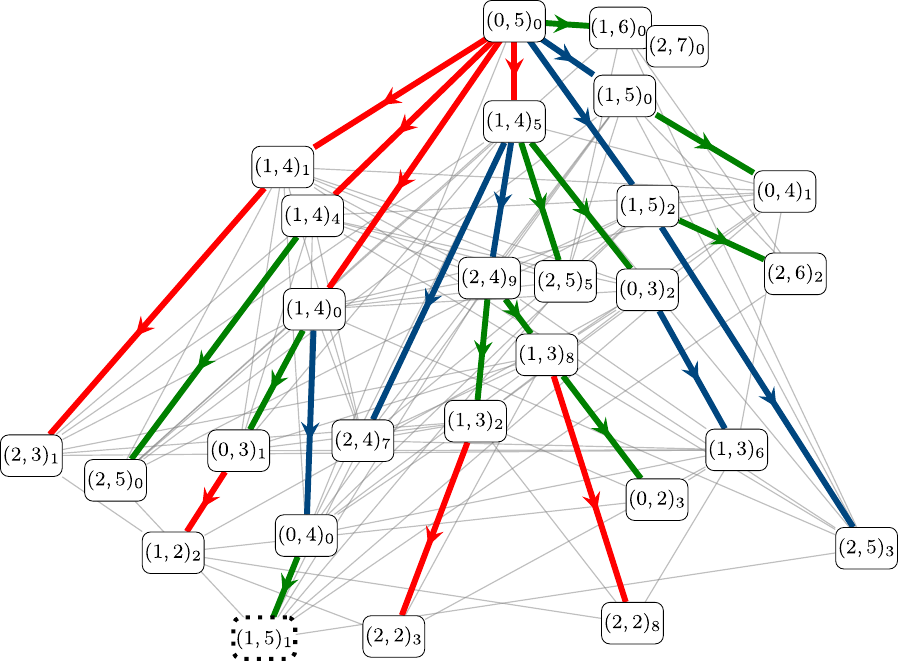}
    \caption{Shortest time paths from the state $(0,5)_0$ to other states in $N=0,1,2$ with unpolarised microwaves. All states shown can be reached with a fidelity of 99.9\% in less than \qty{806}{\micro\second}. The radial distance from $(0,5)_0$ is proportional to the total time required to transfer to a given state. Grey lines indicate dipole-allowed transitions, the superimposed coloured lines forming a tree structure indicate the fastest path to $(0,5)_0$. The colours red, blue, green correspond to driving $\sigma^-,\pi,\sigma^+$ transitions, respectively. The state $(1,5)_1$ is highlighted with a dashed perimeter as it has a dipole-allowed transition from the initial state $(0,5)_0$, but a faster route exists via two intermediate states.}
    \label{fig:graph}
\end{figure}

As an example, we compute $t^{\text{unpol}}_{\pi(a\leftrightarrow b)}$ of $^{87}$Rb$^{133}$Cs at 181.6\,G for all the states with $N=0\rightarrow N_{\text{max}}$ where $N_\text{max}=2$ is the maximum rotational state considered in the calculations.
We store the result in a sparse graph lookup table. Lookups can be implemented in $O(1)$ time, as there is a hashing algorithm that uniquely maps a label pair to a lookup table index. The number of possible transitions scales as $O(N_{\text{max}}^3)$. The most natural graph-theoretic algorithm to run on this graph that has a physical significance is Dijkstra's algorithm~\cite{Dijkstra1959} which in this case computes the shortest time path from a given state to all other states. Figure~\ref{fig:graph} shows the result of running this algorithm starting from the state $(0,5)_0$. The grey lines indicate dipole-allowed transitions and the superimposed bold coloured lines that form a tree structure indicate the the fastest path to $(0,5)_0$. The colours represent the polarisation component of the light that couples to the transition. The radial distance from $(0,5)_0$ is proportional to the total time to get to the given state from $(0,5)_0$. Notably the state $(1,5)_1$, highlighted with a dashed perimeter, according to selection rules is immediately accessible from $(0,5)_0$, however this approach highlights a faster route that exists via two intermediate states.

\subsection{Identifying closed networks of states}

The graph we have described is useful for understanding state transfer, where one microwave $\pi$-pulse is applied at a time. However, implementing synthetic dimensions requires many simultaneously applied driving fields that couple many states~\cite{Sundar2018,Sundar2019}. Similarly, quantum-computing schemes require that the population of some molecules be shelved in off-resonant states, while microwave fields are applied to others to perform gate operations~\cite{Ni2018}. In both cases, each driving field must couple only its target transition and not any other transition that potentially carries a desired population. Therefore, we extend our heuristic approach to such systems.

We consider an arbitrary set of states $S$ in the molecule that are coupled together with microwave fields $M = \{(a\leftrightarrow b) : a,b\in S\}$. For each state, we check for off-resonant excitation by microwaves that do not directly link to that state using a simplified 2-level version of Eq.~\ref{eqn:pulse-time-heuristic} such that
\begin{equation}
2\left< P_2 \right> = \sum_i \frac{1}{1+\kappa_i^2}  \approx  \sum_i \kappa_i^{-2}.
\end{equation}
As with the 3-level model, this allows us to define the duration of a $\pi$-pulse in relation to a desired fidelity
\begin{equation}
t^{\text{sym}}_{\pi(a\leftrightarrow b)} = \max_{c \in S \setminus \{a,b\}}\biggl\{\pi \biggl(\sum_i \frac{1}{\Delta_{ci}^2}\biggl(\frac{d_{ci}}{d_{ab}}\biggr)^2 \biggr)^{1/2}\biggr\} 10^{\eta/2},
\end{equation}
 which we label the \textit{sympathetic time} as it isn't required to meet the fidelity requirement for the transition $a\leftrightarrow b$, but is required to be sympathetic to the overall system, by not affecting the population of the other states we care about during the $\pi$-pulse. In these more complex systems, we call the time required to reach the fidelity on the transition with just a single driving field as the \textit{direct time}, $t^{\text{dir}}_{\pi(a \leftrightarrow b)}$. When comparing transitions in such networks, we have found it useful to set the overall heuristic as the maximum of these two times such that 
\begin{equation}
    t_{\pi(a \leftrightarrow b)} = \max \{t^{\text{sym}}_{\pi(a \leftrightarrow b)} , t^{\text{dir}}_{\pi(a \leftrightarrow b)} \}.
\end{equation}

Unfortunately, the use of this sympathetic concept to our knowledge precludes the use of existing graph theory algorithms. This is because it transforms the value on each edge to be a function of the currently considered states $S$. Therefore, the only apparent solution is to construct all of the possible networks that define the desired structure and evaluate the heuristic for each by brute force. Thankfully our heuristic can be mostly pre-computed into a graph structure and any further in-situ computation of the sympathetic times for each $S$ is easily parallelisable.

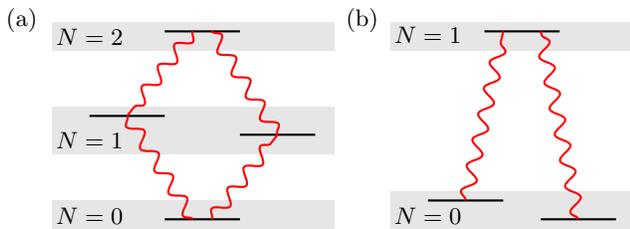
\begin{figure}
    \centering
    \begin{tikzpicture}[scale=2.5]

         \begin{scope}[xshift=+1.7cm]
            \fill[black!10] (-1.9,-0.05) rectangle (-0.6,0.15);
            \fill[black!10] (-1.9,0.9) rectangle (-0.6,1.05);
            
            \draw[black, thick] (-1.1,-0.0) -- (-0.7,0) node[pos=0,left]{$ $};
    
            \draw[black, thick] (-1.7,0.1) -- (-1.3,0.1) node[pos=0,left]{$ $};
    
            \draw[black, thick] (-1.4,1.0) -- (-1.0,1.0) node[pos=0,left]{$ $};

            \draw[-, red, thick, decorate, decoration=snake] (-0.9,0.0) -- (-1.1,1) node[pos=0.5,right]{};
            \draw[-, red, thick, decorate, decoration=snake] (-1.5,0.1) -- (-1.3,1) node[pos=0.5,right]{};
    
            \node[align=left] at (-1.7,0.02) {$N=0$};
            \node[align=left] at (-1.7,0.98) {$N=1$};
        \end{scope}

        \begin{scope}[xshift=-1.8cm]
    
            \fill[black!10] (-0.2,-0.05) rectangle (1.3,0.1);
    
            \fill[black!10] (-0.2,0.9) rectangle (1.3,1.05);
    
            \fill[black!10] (-0.2,0.6) rectangle (1.3,0.35);
            
            \draw[black, thick] (0.4,-0.0) -- (0.8,0) node[pos=0,left]{$ $};
            \draw[black, thick] (0.0,0.55) -- (0.4,0.55) node[pos=0,left]{$ $};
    
            \draw[black, thick] (0.8,0.45) -- (1.2,0.45) node[pos=0,left]{$ $};
    
            \draw[black, thick] (0.4,1.0) -- (0.8,1.0) node[pos=0,left]{$ $};
    
            \node[align=left] at (0.,0.97) {$N=2$};
            \node[align=left] at (0.,0.42) {$N=1$};
    
            \node[align=left] at (0.,0.02) {$N=0$};
    
            \draw[-, red, thick, decorate, decoration=snake] (0.55,0.0) -- (0.2,0.55) node[pos=0.5,right]{};
            \draw[-, red, thick, decorate, decoration=snake] (0.65,0.0) -- (1.0,0.45) node[pos=0.5,right]{};
            \draw[-, red, thick, decorate, decoration=snake] (0.55,1.0) -- (0.2,0.55) node[pos=0.5,right]{};
            \draw[-, red, thick, decorate, decoration=snake] (0.65,1.0) -- (1.0,0.45) node[pos=0.5,right]{};
        \end{scope}

        \node[align=left] at (-2.16,1.06) {(a)};
        \node[align=left] at (-0.35,1.06) {(b)};

    \end{tikzpicture}
    \caption{Network geometries we consider as examples. (a)~4-level loop for the implementation of a 1D synthetic dimension with periodic boundary conditions. (b)~3-level $\Lambda$ system that is applicable to the iSWAP protocol outlined in~\cite{Ni2018}. Here, the two $N=0$ states form a storage qubit, and an $N=0,1$ pair form an interacting qubit. }
    \label{fig:geometries}
\end{figure}

We use this approach to find optimal sets of states in $^{87}$Rb$^{133}
$Cs across a magnetic field range of $1\,\mathrm{G}$ to $500\,\mathrm{G}$ for two different experiments; 4-level synthetic dimension with periodic boundary conditions, and a 3-level system that may be used for the iSWAP protocol proposed by Ni~${et~al.}$~\cite{Ni2018}, and recently demonstrated in experiments using single NaCs atoms in optical tweezers~\cite{Picard2025}. These two geometries are shown schematically in Fig.~\ref{fig:geometries}. In either case, we fix the initial state of the molecules to be either $(0,4)_1$ or $(0,5)_0$, however these states do not have to be one of the optimal states. Rather, we evaluate transfers within the structures as $t_\text{structure} \equiv \max_{(a\leftrightarrow b)\in M}{\{t^{\text{unpol}}_{\pi(a\leftrightarrow b)}\}}$, based on the smallest Rabi frequency within the structure for a given fidelity assuming unpolarised microwaves. We also consider the total time $t_\text{travel}$ for the transfer of the molecules to the nearest state in that structure with that same fidelity and unpolarised microwaves. To rate the set of states, we then combine these quantities via a ranking metric $R_t$ as
\begin{equation}
R_t = [f \,t_\text{travel} + (1-f)|M|\,t_\text{structure}]^{-1}.
\end{equation}
The relative weighting~$f$ given to the structure time and travel time can have a significant impact on the optimum sets of states; we choose to use a value of $f=0.2$ throughout this work. This weighting allows the travel time to have a reasonable effect on the optimisation, which is useful for our demonstration. 

\begin{figure}[t]
    \centering
    \includegraphics{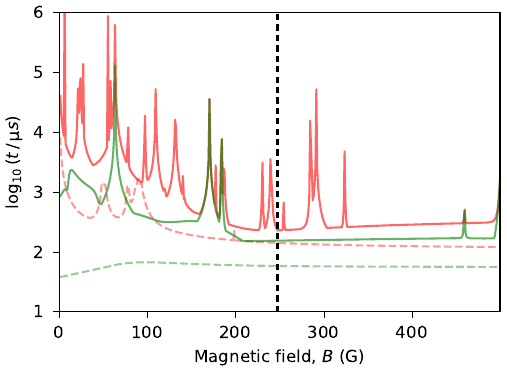}
    \caption{Variation of timescales to meet 99.9\% fidelity for the optimum set of states $\{(0,2)_3,(1,3)_8,(2,2)_8,(1,2)_4\}$ to implement a 4-level synthetic dimension with periodic boundary conditions as rated by the ranking metric $R_t$. The solid lines represent $t_\text{structure}$ for polarised (unpolarised) microwaves in green (red). The dashed lines represent $t_\text{travel}$ with the same respective polarisation-colour mapping. The vertical dashed line at \qty{247}{G} is the value of the magnetic field $B$ that maximises $R_t$.}
    \label{fig:4-loop-heuristic}
\end{figure}

For the 4-level synthetic dimension, the set of states with the highest rating found are $\{(0,2)_3,(1,3)_8,(2,2)_8,(1,2)_4\}$ at a field of $247\,\mathrm{G}$. We quote the heuristic times with respect to a desired fidelity of $F=99.9\%$~($\eta=3$). The loop can be entered with the transition $(0,4)_1\rightarrow (1,3)_8$ with a travel time $t^{\text{pol/unpol}}_\text{travel} =\qty{59}{\micro\second}\,\,/\,\,\qty{140}{\micro\second}$, and the maximum $t^{\text{pol/unpol}}_{\pi(a\leftrightarrow b)}$ within the structure is  $\qty{160}{\micro\second}\,\,/\,\,\qty{230}{\micro\second}$  for completely polarised or unpolarised microwaves respectively. To better understand the sensitivity of this choice to magnetic field, in Fig.~\ref{fig:4-loop-heuristic} we show the computed heuristic $t_\text{structure}$ for polarised and unpolarised microwaves in the solid green and red lines respectively against magnetic field. The spikes in such plots correspond to degeneracies in energy between intended and unintended transitions, or zeros of an intended-transition dipole moment. The similarly coloured dashed lines represent the $t_\text{travel}$ heuristic. The vertical black dashed line indicates the optimum field of $247\,\mathrm{G}$ that maximises the metric $R_t$.

In the previous example, our ranking solely considered the negative effects of off-resonant states, however other considerations can be simultaneously incorporated in the optimisation routine.
For example, in the 3-level quantum computing protocol, it is important that the three states possess similar magnetic moments such that they become insensitive to experimental magnetic field noise~\cite{Gregory2021,Lin2022}. The infidelity due to decoherence over a time $t$ associated with magnetic field noise can be approximated by
${(\Delta\mu)(\Delta B)t}/{\hbar} \equiv 10^{-\eta}$. Here, $\Delta\mu$ is the largest difference between magnetic dipole moments in the system, and $\Delta B$ is the magnitude of the magnetic field noise. To evaluate the sensitivity of a pair of states, we rearrange this equation for the maximum magnetic field noise allowed to match the fidelity achieved by the microwave transfer with $t=t_\pi$ to find
\begin{equation}
    \Delta B_\text{max } = \frac{2\hbar}{\pi (\Delta\mu)}\biggl({\sum_i\frac{D_i^2}{\Delta_i^2}}\biggr)^{-1/2}\,10^{-3\eta/2}.
\end{equation}
To incorporate this metric into our ranking of structures, we use a similar quantity but use $t=t_\text{structure}$ instead. Our overall ranking metric is then given by $R_t (\Delta B_\text{max})^e$ where $e$ is the weighting given to the maximum magnetic field noise. Here, we manually tuned the value of $e$ until we found optimum sets of states that were compatible with the $\sim10$\,mG magnetic field noise achieved in experiments in our own group~\cite{Hepworth2025}. Following this strategy, we fixed $e=1/3$. 

\begin{figure}[t]
    \centering
    \includegraphics{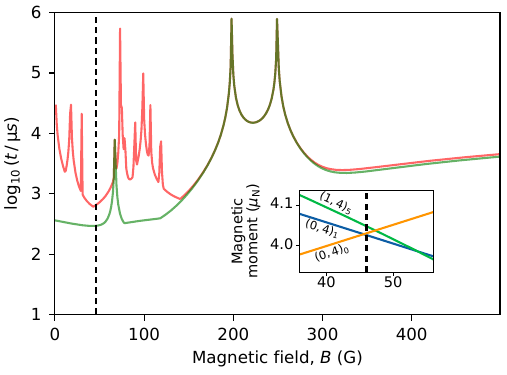}
    \caption{Variation of timescales to meet 99.9\% fidelity for the optimum set of states $\{(0,4)_1,(1,4)_5,(0,4)_0\}$ to implement a robust storage qubit as rated by the ranking metric $R_t (\Delta B_\text{max})^{1/3}$. Lines in the main plot are as in Fig.~\ref{fig:4-loop-heuristic}. The additional inset shows the coincident magnetic dipole moments of the three states around the optimum ranking field of~\qty{47}{G}.}
    \label{fig:N0-qubit}
\end{figure}

The set of states with the highest ranking found for the 3-level quantum-computing protocol are $\{(0,4)_1,(1,4)_5,(0,4)_0\}$ at a field of \qty{47}{G}. The maximum $t^{\text{pol/unpol}}_{\pi(a\leftrightarrow b)}$ within the structure is $\qty{300}{\micro\second}\,\,/\,\, \qty{670}{\micro\second}$ for completely polarised or unpolarised microwaves, respectively. Here, $(0,4)_1$ is a state we can reach immediately from STIRAP, so $t^\text{pol/unpol}_\text{travel}=0$. Similar to the aforementioned 4-level synthetic dimension, Fig.~\ref{fig:N0-qubit} shows the heuristic $t_\text{structure}$ against magnetic field $B$, however here we also show in the inset the magnetic moments of the three states. The two hyperfine qubit states $(0,4)_1$ and $(0,4)_0$ cross at this field, while the adjoining state $(1,4)_5$ coincidently has a similar magnetic dipole moment. The acceptable magnetic field noise for this state combination is~$\Delta B^\text{pol/unpol}_\text{max}$ is \qty{230}{\milli G} / \qty{100}{\milli G}.

The exact parameters of the optimisation may be varied depending on the needs of a given experiment. For the two examples given we have restricted the rotational states to $N=0$~and~$1$ as shown in Fig.~\ref{fig:geometries}. However, a broader search could also be conducted without this restriction if suitable states were not found. Moreover, our ranking approach can be easily modified to  search for sets of states that satisfy other requirements. For example, it may be useful to choose from all sets of states that are degenerate in energy to realise spin-orbital dynamics as proposed in~\cite{Syzranov2014}. While we have used a bialkali as an example, our methods are generally applicable to any molecule with dense hyperfine structure, or indeed can be applied to any quantum system comprised of dense manifolds of states where transitions can only be driven between states occupying different manifolds; similar structure for example also exists in the Rydberg states of atoms where graph-based paradigms have also been found useful~\cite{Miller2023}.

In conclusion, we have shown how to determine optimal pathways and isolated networks of states in bialkali molecules. Our approach relies on a simple heuristic, based on breaking the many transitions in the molecule down to sets of 3 levels at a time, to estimate the maximum speed with which the transfer between any pair of states can be performed. This allows us to construct a graph of the molecular transitions that shows the relative distance between any pair of states, and to use established algorithms from graph theory to quickly search the graph for optimal routes. Searching for isolated networks of states is more computationally intensive, but still relatively fast based on our approach. As an example, we have presented the optimum sets of states in the RbCs molecule  to form a closed 4-level loop representing a synthetic dimension with periodic boundary conditions and the optimal sets of states to implement an iSWAP quantum entangling gate between molecular hyperfine qubits. Our work presents a flexible approach to choosing optimal state couplings that is immediately applicable to all experiments with ultracold molecules. 

\section{Code Availability}

All calculations of the rotational and hyperfine structure of RbCs are performed using the open-source Diatomic-Py code that we have previously published in~\cite{Blackmore2023}, with the newest version \href{https://github.com/durham-qlm/diatomic-py}{available here}. The code to generate these figures is available on GitHub at \url{https://github.com/durham-qlm/diatomic-networks}, and is written to be adaptable to many bialkali systems.

\section{Author Contributions}
All authors contributed to the conceptualisation of this project. TRH developed the codes and performed all calculations with supervision by SLC and PDG. The manuscript was first written by TRH and PDG, and was then reviewed by all authors.

\section{Acknowledgements}

We acknowledge support from the UK Engineering and Physical Sciences Research Council (EPSRC) Grants EP/P01058X/1, EP/W00299X/1 and UKRI2226 funded through the Programme Grant Scheme, UK Research and Innovation (UKRI) Frontier Research Grant EP/X023354/1, the Royal Society, and Durham University. PDG is supported by a Royal Society University Research Fellowship URF/R1/231274 and Royal Society research grant RG/R1/241149.

\bibliography{references}

\clearpage
\newpage
\onecolumngrid

\section*{Supplemental Material}

\subsection*{Analytic solution to the 3-level time evolution}

The Hamiltonian that describes our 3-level model for off-resonant excitation in the basis $\{\ket{0},\ket{1},\ket{2}\}$ is
\begin{equation}
    \tilde{H} =
    \frac{\hbar \Omega}{2}
    \begin{pmatrix}
        0 & 1 & D\\
        1 & -2f\kappa & 0\\
        D & 0 & 2(1-f)\kappa
    \end{pmatrix}.
\end{equation}
Solving the matrix part as an eigenvalue equation,
\begin{equation}\label{eqn:chareq}
    \begin{vmatrix}
        -\mu & 1 & D\\
        1 & -\mu-2f\kappa & 0\\
        D & 0 & -\mu+2(1-f)\kappa
    \end{vmatrix}
    =
    \mu^3 - 2 (1-2f) \kappa \mu^2 - (1 + D^2 +4(1-f)f\kappa^2) \mu - 2(-1+f+f D^2)\kappa = 0,
\end{equation}
one finds 3 roots $\mu_1,\mu_2,\mu_3$ proportional to the eigenvalues. This lets us write the Hamiltonian as $\tilde{H} = PDP^{-1}$ where
\begin{equation}
    D = 
    \frac{\hbar \Omega}{2}
    \begin{pmatrix}
        \mu_1 & 0 & 0\\
        0 & \mu_2 & 0\\
        0 & 0 & \mu_3
    \end{pmatrix},
    \quad
\end{equation}
\begin{equation}
    P = 
    \begin{pmatrix}
        \mu_1-2\kappa (1-f) & \mu_2-2\kappa (1-f) & \mu_3-2\kappa (1-f) \\
        \mu_1 (\mu_1-2\kappa(1-f))-D^2 & \mu_2 (\mu_2-2\kappa(1-f))-D^2 & \mu_3 (\mu_3-2\kappa(1-f))-D^2\\
        \Gamma & \Gamma & \Gamma
    \end{pmatrix}.
\end{equation}
We can then use this diagonal form to propagate the initial state vector $\psi(t=0) = (\sqrt{p},0,\sqrt{1-p})^T$, that is, one with a population $p$ in the ground state, and population $1-p$ in the off-resonant state with
\begin{equation}
    \psi (t) = e^{-i\tilde {H} t/ \hbar}\psi (0) =     P e^{i D t/ \hbar} P^{-1} \psi (0).
\end{equation}
Due to the solution's invariance to the exchange of roots $\mu_i\leftrightarrow\mu_j$, they can become unwieldy, so we will write them in summation notation by defining
\begin{equation}
    R_{jk}\coloneqq (\mu_j-\mu_k) \left({D} \sqrt{p} \, ({\mu_j}+{\mu_k}- 2 (1-f) \kappa)-\sqrt{1-p} \,({\mu_j}-2 (1-f) \kappa) (\mu_k-2 (1-f) \kappa)-D^2 \sqrt{1-p}\right),
\end{equation}
\begin{equation}
    A_i \coloneqq \sum_{j,k} \frac{1}{2}\epsilon_{ijk}      (\mu_i-2(1-f)\kappa)      R_{jk},
\end{equation}
\begin{equation}
    B_i \coloneqq \sum_{j,k} \frac{1}{2}\epsilon_{ijk}       (\mu_i(\mu_i-2(1-f)\kappa)-D^2)          R_{jk},
\end{equation}
\begin{equation}
    C_i \coloneqq \sum_{j,k} \frac{1}{2}\epsilon_{ijk}           D               R_{jk}.
\end{equation}
We can then write the time-evolved wave function as
\begin{equation}
    \psi(\tau) = P e^{iDt/\hbar} P^{-1} \psi(0)
    = \frac{1}{D(\mu_1-\mu_2)(\mu_3-\mu_1)(\mu_2-\mu_3)}
    \begin{pmatrix}
        \sum_{i} A_{i} e^{-i\mu_{i}\tau/2}\\
        \sum_{i} B_{i} e^{-i\mu_{i}\tau/2}\\
        \sum_{i} C_{i} e^{-i\mu_{i}\tau/2}
    \end{pmatrix}.
\end{equation}
Using the identity
\begin{equation}
\begin{aligned}
    \left(\sum_i A_i e^{-i\theta_i}\right)\left(\sum_j A_j e^{i\theta_j}\right)
    = \sum_{i,j} A_i A_j \cos(\theta_i-\theta_j)
    = \sum_{i} A_i^2 + 2\sum_{i>j} A_i A_j \cos(\theta_i-\theta_j),
\end{aligned}
\end{equation}
we can find the state probabilities over time as
\begin{equation}\label{eqn:state-probs}
    P_i(\tau) = 
    \frac{1}{(D(\mu_1-\mu_2)(\mu_3-\mu_1)(\mu_2-\mu_3))^2}
    \begin{pmatrix}
        \sum_{i} A_i^2 + 2\sum_{i > j} A_i A_j \cos((\mu_i-\mu_j) \tau/2)\\
        \sum_{i} B_i^2 + 2\sum_{i > j} B_i B_j \cos((\mu_i-\mu_j) \tau/2)\\
        \sum_{i} C_i^2 + 2\sum_{i > j} C_i C_j \cos((\mu_i-\mu_j) \tau/2)
    \end{pmatrix}.
\end{equation}

The average quantity of interest $2\langle P_2 \rangle_t$ (for $p=1$) is given by
\begin{multline}
 2\langle P_2 \rangle_t \big\rvert_{p=1} = \frac{2\sum_{i} B_i^2}{(D(\mu_1-\mu_2)(\mu_3-\mu_1)(\mu_2-\mu_3))^2}\bigg\rvert_{p=1}=\\
\frac{-D^2 \left(4 (f-4) \kappa ^2-2\right)+\left(4 (f-1) \kappa ^2+1\right)^2+D^4}{\left(4 (f-1) \kappa ^2+1\right)^2 \left(f^2 \kappa ^2+1\right)+D^4 \left(((f-10) f+1) \kappa ^2+3\right)+D^2 \left(2 \kappa ^2 \left(f \left(-4 ((f-4) f+1) \kappa ^2+f-1\right)+10\right)+3\right)+D^6}.    
\end{multline}
This simplification can by performed by a reduction over the symmetric polynomials of the roots $\mu_i$ with the characteristic polynomial (Eq.~\ref{eqn:chareq}) coefficients. By substituting $f=(\sqrt{1+(D^2/\kappa^2)}-1)/2$ we get what we define to be $2\langle P_2 \rangle^\text{comp}_t$ and this is what we plot in Fig.~\ref{fig:3-level-phase-intended} of the main text.

\end{document}